\documentclass[runningheads]{llncs}
\usepackage[utf8]{inputenc} 
\usepackage[T1]{fontenc}    
\usepackage[colorlinks]{hyperref}       
\usepackage{url}            
\usepackage{booktabs}       
\usepackage{amsfonts}       
\usepackage{nicefrac}       
\usepackage{microtype}      
\usepackage{graphicx}
\usepackage{natbib}
\usepackage{doi}
\usepackage{graphicx}
\usepackage{svg}

\usepackage{todonotes}

\newcommand{\ignore}[1]{}

\title{Idiosyncratic properties of Australian STV election counting\thanks{This work was supported by the Australian Research Council (Discovery Project DP220101012, OPTIMA ITTC IC200100009).}}

\author{ \href{https://orcid.org/0000-0001-6277-2442}{\includegraphics[scale=0.06]{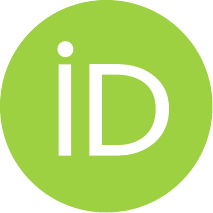} \hspace{1mm} Andrew Conway}\thanks{\url{https://www.andrewconway.org}. Andrew Conway is a member of the Secular Party. This paper has nothing to do with the Secular Party.} \\
 	Melbourne, Australia \\
 	\and
  \href{https://orcid.org/0000-0002-0459-9917}{\includegraphics[scale=0.06]{orcid.pdf}\hspace{1mm}Michelle Blom}\\
  School of Computing and Information Systems \\
  The University of Melbourne 
}

\author{
Andrew Conway\inst{1}
\thanks{Andrew Conway is a member of the Secular Party. This paper has nothing to do with the Secular Party.}
\and
Michelle Blom\inst{2}
\and
Alexander Ek\inst{3}
\and
Peter J. Stuckey\inst{4}
\and
Vanessa J. Teague\inst{5}
\and
Damjan Vukcevic\inst{3}
}
\authorrunning{Conway, Blom, Ek, Stuckey, Teague, and Vukcevic}

\institute{E-Vote-ID 2024}
\institute{
\url{https://www.andrewconway.org}, Melbourne, Australia
\and
\email{michelle.blom@unimelb.edu.au}, School of Computing and Information Systems, University of Melboure,
Parkville, Australia
\and
Department of Econometrics and Business Statistics, Monash University, Clayton,
Australia
\and
Department of Data Science and AI, Monash University, Clayton, Australia
\and
Australian National University and
Thinking Cybersecurity Pty Ltd. 
}

\hypersetup{
pdftitle={Idiosyncratic properties of Australian STV election counting},
pdfsubject={q-bio.NC, q-bio.QM, TODO FIX},
pdfauthor={Andrew Conway, others},
pdfkeywords={STV, Elections},
}

\begin{document}
\maketitle

\begin{abstract}
 Single Transferable Vote (STV) counting, used in several jurisdictions in Australia, is a system for choosing multiple election winners
 given voters' preferences over candidates.
 There are a variety of different versions of STV legislated and/or applied across Australia. This paper shows some of the unintuitive properties of some of these systems.
\end{abstract}

Preferential voting has been a feature of the Australian electoral landscape since the early 1900s. In a preferential (or ranked-choice) election, voters cast their vote by recording a preference list over the available candidates. In Australia, two main election systems are used: Instant-Runoff Voting (IRV); and Single Transferable Vote (STV). Both are preferential. IRV is used to elect a single winner. STV elects multiple candidates to fill a number of seats. The IRV counting algorithm has been implemented more or less uniformly across Australian jurisdictions. STV tabulation, in contrast, has attracted a variety of quite different interpretations. Typically, STV tabulation proceeds as follows:
\begin{itemize}
    \item Step 1: Each vote gets distributed to the highest-ranked candidate.
    \item Step 2: Determine a quota, typically being slightly more than the number of valid votes
                  divided by one more than the number of vacancies (positions to fill). The idea is that it should
                  be impossible for more candidates than the number of vacancies to simultaneously reach a quota.

    \item Step 3: If a candidate has at least a quota, they are considered elected.
    \item Step 4: If enough candidates are elected, stop.
    \item Step 5: If the number of vacancies not already filled equals the number of \emph{continuing} candidates 
                  (those neither elected nor excluded), declare them elected and stop.
    \item Step 6: If an elected candidate had more than a quota, distribute the surplus votes (those over a quota)
                  to the next continuing candidate (neither elected nor excluded), and go back to step 3.
    \item Step 7: Exclude the candidate with the fewest votes. Distribute their votes
                  to the first continuing candidate on the preference list.
    \item Step 8: Go to step 3.
\end{itemize}

Even ignoring irrelevant variations that stop the process early, or tie resolutions,
there are lots of ambiguities in this approach. 
The biggest is in surplus distribution, step 6. Which votes are considered the
surplus votes? Should they be just chosen from the most recently received
votes (ACT, \emph{last parcel}) or perhaps the last few sets of most recently received votes (NSW)? 
Should they be chosen randomly (NSW)? Should they all be
distributed but each only count as a fraction of a vote (this fraction is known as a \emph{transfer value})? 
If so, and some of the votes for the candidate being distributed had different transfer values, should they
be distributed with different transfer values (most states) or the same ones (Federal and Victoria)? 
Should the resulting fractional votes
be computed exactly (no one), or rounded to integers (almost everyone, rounded down), 
or to six decimal places (ACT 2020 and later, rounded down)? If there are multiple 
outgoing transfer values, should they be transferred in different batches and rounded separately (everyone)? 
Should whether or not a vote is exhausted affect the transfer value given to other 
votes (NSW local, ACT yes; Federal, Victoria, WA no)?

Different answers to any of the above questions can produce different winners. \citet{gore2016simulating} consider how the STV implementation used by the Australian Capital Territory, containing simplifications designed to aid hand-counting, including the use of \textit{last parcel} surplus transfer, may lead to unintuitive results. \citet{schurmann2018rounding} considers how the use of rounding during tabulation of proportional representation elections can change their outcomes.

This paper presents three properties that should be intuitive for a preferential election system, but that do not hold for at least one implementation of STV in Australia. That: a candidate tally cannot become negative (Fallacy 1); votes cannot increase in value during tabulation (Fallacy 2); and taking votes away from a losing candidate cannot make that candidate win (Fallacy 3).

An open source program ConcreteSTV\footnote{\url{https://github.com/AndrewConway/ConcreteSTV}}
has implemented several jurisdictions' legislation and/or used implementations, matching known bugs in various
years, allowing a comparison of how the different implementations compare in practice. The examples described in this paper can be replicated with ConcreteSTV.

\subsubsection*{Fallacy 1: Candidates cannot get a negative tally.}\label{sec:NegativeTally}

A candidate with a negative tally seems absurd, but could occur with a negative transfer value. This appears  possible in the NSW local government legislation [7(4)(a)]\footnote{Schedule 5 
of the N.S.W. Local Government (General) Regulation 2005, \url{https://legislation.nsw.gov.au/view/whole/html/2020-10-27/sl-2005-0487\#sch.5}}: \textit{the number of surplus votes is to be divided by the number of votes received by the elected candidate (excluding the aggregate value of any exhausted votes). The surplus fraction is equal to the resulting fraction or (if the fraction exceeds 1) to 1}.

Consider a seated candidate $c$ with tally $V$ in an election with quota $Q$. Their surplus is $V-Q$ and surplus fraction $(V-Q)/(V-E)$ with $E$ denoting the aggregated value of exhausted votes. The legislation attempts to assign $E$ to the $Q$ votes left behind, increasing the surplus fraction of the distributed votes. If $E\leq Q$, then the outgoing value
of the non-exhausted votes being distributed will be $V-Q$ as desired. If $Q<E<V$ then there are more exhausted votes
than are needed to fill the quota, and the most reasonable thing to do is to distribute the non-exhausted votes with the
transfer values they came to the current candidate with. This is accomplished by the last part of 7(4)(a) as the formula
would indeed give a fraction exceeding 1 and so the surplus fraction is set to one.

However, if we look carefully at how $V$ and $E$ are defined, a problem
crops up. $V$ is the tally that the candidate had when elected, which is the
sum over each count of the number of ballots received at that count times their associated transfer value, {\em rounded
down to an integer at each count}. The ``aggregate value of any exhausted votes'' is not clearly defined. The legislation doesn't give any support for any definition. The most obvious, which
matches what the NSW Electoral Commission actually did, is as the sum of all the transfer values of all the exhausted 
votes.  The reason this is a problem is that $V$ is the sum of many values, with a potentially large amount of
rounding down, and $E$ is the sum of a subset of these values, {\em with no rounding}. This means that it is
possible for $E$ to be larger than $V$. In this case the denominator of the surplus fraction is
negative, which means that the fraction itself is negative.

A specific (artificial) example in which this occurs is available at:\url{https://vote.andrewconway.org/Oddities/NSW%20LGE/NegativeTally/Recount.html}.

\subsubsection*{Fallacy 2: Votes cannot increase in value during tabulation.}\label{sec:IncreaseVotes}

The Australian Senate's variation of STV computes the transfer value for ballots in an elected candidate $c$'s tally by dividing their surplus by the \textit{total number of ballots} in their tally, \textit{rather than} the total value of these ballots. Each ballot in the tally is assigned \textit{the same transfer value} when being distributed to continuing candidates. The seated candidate may, in previous rounds of tabulation, have received votes from prior surplus transfers. These votes, which came to $c$ at value $v$, may very well now have a value $v' > v$, as they have been reweighted. This means that the overall contribution of these ballots to the tabulation will be a value of more than 1.   

This phenomenon occurs frequently in the tabulation of Australian Senate
elections, and can be observed in around a third of recent such elections. For
instance, in the 2016 Senate election for Tasmania, counted with the AEC2016
ruleset, J. Dunian (the 3rd listed Liberal candidate) has a surplus distribution
with transfer value 0.144792 in count 8. This means that all ballots leaving J.
Dunian are assigned that value.  In count 5, J Dunian received 8 ballots from
the distribution of Jacqui Lambie, each with value 0.0640023. These ballots
continued, increasing in value to 0.144792 on the election of J. Dunian. This example can be examined at \url{https://vote.andrewconway.org/Federal%20Senate/2016/TAS/Recount.html}.

\subsubsection*{Fallacy 3: Taking votes away from a loser cannot make them win.}\label{sec:TakingVotesAwayFromLoser}

A losing candidate who complained that they would have won if they had fewer votes
sounds like a paranoid fantasy, but this is actually a plausible scenario. Some heuristics have shown that it did occur 
in the 2013 Australian Senate election for Victoria, where Ricky Muir won a seat. 
Ricky Muir was very close to elimination on count 223, where the 3 candidates with lowest tallies were: Joe Zammit on 16,404; Vickie Janson on 16,683; and Ricky Muir on 18,994. As the lowest, Joe Zammit gets excluded and 15,909 of the ballots pass to Ricky Muir, who suddenly
has a much stronger position. Over subsequent rounds, Ricky Muir gained a substantial number of votes from eliminated candidates, resulting in him eventually winning the last seat over Helen Kroger who, on count 223, had 391,818 votes. However, if Helen Kroger  were to give\footnote{
That is, persuade some people who voted for her ahead of Joe Zammit to give him the
preference they originally gave her, and her the inferior preference they  gave him.} 2595 votes to Joe Zammit, then Joe Zammit would no longer
have the smallest tally at count 223; Vickie Janson would. Very few of her votes would pass to Ricky Muir,
and then Ricky Muir would still be behind Joe Zammit, and would be next eliminated. With Ricky
Muir out of the race, Helen Kroger would end up winning with an apparent margin of over 87,000 votes.

This example is available at: \url{https://vote.andrewconway.org/Federal%20Senate/2013/VIC/Recount.html}.

This non-monotonicity property is well known for STV, and indeed is a consequence of (the failure of) Arrow's \emph{Independence of Irrelevant Alternatives} condition, but it is not generally appreciated how often it arises in practice.

\section*{Conclusion}
There are many properties we would expect from a sensible election system. Unfortunately for various reasons the election systems we legislate often do not take into account possible strange edge cases. Ideally legislators should talk to mathematicians or computer scientists to ensure that their legislation is unambiguous and avoids problematic edge cases.

\bibliographystyle{unsrtnat}
\renewcommand*{\bibfont}{\interlinepenalty 10000\relax}
\renewcommand{\bibsection}{\subsection*{Bibliography}}
\bibliography{references} 

\begin{thebibliography}{2}
\providecommand{\natexlab}[1]{#1}
\providecommand{\url}[1]{\texttt{#1}}
\expandafter\ifx\csname urlstyle\endcsname\relax
  \providecommand{\doi}[1]{doi: #1}\else
  \providecommand{\doi}{doi: \begingroup \urlstyle{rm}\Url}\fi

\bibitem[Gor{\'e} and Lebedeva(2016)]{gore2016simulating}
Rajeev Gor{\'e} and Ekaterina Lebedeva.
\newblock Simulating {STV} hand-counting by computers considered harmful:
  {ACT}.
\newblock In \emph{International Joint Conference on Electronic Voting}, pages
  144--163. Springer, 2016.

\bibitem[Sch{\"u}rmann(2018)]{schurmann2018rounding}
Carsten Sch{\"u}rmann.
\newblock Rounding considered harmful.
\newblock In \emph{Electronic Voting: Third International Joint Conference,
  E-Vote-ID 2018, Bregenz, Austria, October 2-5, 2018, Proceedings 3}, pages
  189--202. Springer, 2018.

\end{thebibliography}

\end{document}